%% file: LHCb-PAPER-2016-022.tex
\pdfoutput=1

\documentclass[12pt,a4paper]{article}

\usepackage{ifthen} 
\newboolean{pdflatex}
\setboolean{pdflatex}{true} 

\newboolean{articletitles}
\setboolean{articletitles}{true} 

\newboolean{uprightparticles}
\setboolean{uprightparticles}{false} 

\newboolean{inbibliography}
\setboolean{inbibliography}{false} 

\input{preamble}
\usepackage{longtable} 
\input{bhhh-symbols-def}

\begin{document}

\renewcommand{\thefootnote}{\fnsymbol{footnote}}
\setcounter{footnote}{1}

\input{title-LHCb-PAPER}


\renewcommand{\thefootnote}{\arabic{footnote}}
\setcounter{footnote}{0}



\pagestyle{plain} 
\setcounter{page}{1}
\pagenumbering{arabic}


%

\input{introduction}

\input{detector_short}

\input{selection_short}
\input{fits}

\input{eff_short}
\input{results}

\input{acknowledgements}


\input{LHCb-PAPER-2016-022.bbl}

\newpage
\input{LHCb_Authorship_flat_01-Jun-2016.tex}
\end{document}

%% file: preamble.tex
\usepackage{microtype}
\usepackage{lineno}  
\usepackage{xspace} 
\usepackage{caption} 

\usepackage{graphicx}  
\usepackage{color}
\usepackage{colortbl}
\graphicspath{{./figs/}} 

\usepackage{amsmath} 
\usepackage{amssymb}
\usepackage{amsfonts}
\usepackage{upgreek} 

\newcommand*\patchAmsMathEnvironmentForLineno[1]{%
\expandafter\let\csname old#1\expandafter\endcsname\csname #1\endcsname
\expandafter\let\csname oldend#1\expandafter\endcsname\csname
end#1\endcsname
 \renewenvironment{#1}%
   {\linenomath\csname old#1\endcsname}%
   {\csname oldend#1\endcsname\endlinenomath}%
}
\newcommand*\patchBothAmsMathEnvironmentsForLineno[1]{%
  \patchAmsMathEnvironmentForLineno{#1}%
  \patchAmsMathEnvironmentForLineno{#1*}%
}
\AtBeginDocument{%
\patchBothAmsMathEnvironmentsForLineno{equation}%
\patchBothAmsMathEnvironmentsForLineno{align}%
\patchBothAmsMathEnvironmentsForLineno{flalign}%
\patchBothAmsMathEnvironmentsForLineno{alignat}%
\patchBothAmsMathEnvironmentsForLineno{gather}%
\patchBothAmsMathEnvironmentsForLineno{multline}%
\patchBothAmsMathEnvironmentsForLineno{eqnarray}%
}

\input{lhcb-symbols-def} 

\usepackage{cite} 
\usepackage{mciteplus}
\usepackage{hyperref}    
\usepackage[all]{hypcap} 

%% file: lhcb-symbols-def.tex
\def\lhcb {\mbox{LHCb}\xspace}

\ifthenelse{\boolean{uprightparticles}}%
{

 \def\Ppi         {\ensuremath{\uppi}\xspace}

 \def\Pchi        {\ensuremath{\upchi}\xspace}                 
 \def\Ppsi        {\ensuremath{\uppsi}\xspace}

 \def\PDelta      {\ensuremath{\Delta}\xspace}                 
 \def\PXi      {\ensuremath{\Xi}\xspace}                 
 \def\PLambda      {\ensuremath{\Lambda}\xspace}                 
 \def\PSigma      {\ensuremath{\Sigma}\xspace}                 
 \def\POmega      {\ensuremath{\Omega}\xspace}                 
 \def\PUpsilon      {\ensuremath{\Upsilon}\xspace}                 
 
 \def\PB      {\ensuremath{\mathrm{B}}\xspace}                 
                  
 \def\PD      {\ensuremath{\mathrm{D}}\xspace}

 \def\PJ      {\ensuremath{\mathrm{J}}\xspace}                 
 \def\PK      {\ensuremath{\mathrm{K}}\xspace}

 \def\Pb      {\ensuremath{\mathrm{b}}\xspace}                 
 \def\Pc      {\ensuremath{\mathrm{c}}\xspace}

 \def\Pi      {\ensuremath{\mathrm{i}}\xspace}

 \def\Ps      {\ensuremath{\mathrm{s}}\xspace}

}
{

 \def\Ppi         {\ensuremath{\pi}\xspace}

 \def\Pchi        {\ensuremath{\chi}\xspace}                 
 \def\Ppsi        {\ensuremath{\psi}\xspace}                 
                  
 \mathchardef\PDelta="7101
 \mathchardef\PXi="7104
 \mathchardef\PLambda="7103
 \mathchardef\PSigma="7106
 \mathchardef\POmega="710A
 \mathchardef\PUpsilon="7107
                  
 \def\PB      {\ensuremath{B}\xspace}                 
                  
 \def\PD      {\ensuremath{D}\xspace}

 \def\PJ      {\ensuremath{J}\xspace}                 
 \def\PK      {\ensuremath{K}\xspace}

 \def\Pb      {\ensuremath{b}\xspace}                 
 \def\Pc      {\ensuremath{c}\xspace}

 \def\Pi      {\ensuremath{i}\xspace}

 \def\Ps      {\ensuremath{s}\xspace}

}

\makeatletter
\ifcase \@ptsize \relax
  \newcommand{\miniscule}{\@setfontsize\miniscule{4}{5}}
\or
  \newcommand{\miniscule}{\@setfontsize\miniscule{5}{6}}
\or
  \newcommand{\miniscule}{\@setfontsize\miniscule{5}{6}}
\fi
\makeatother

\DeclareRobustCommand{\optbar}[1]{\shortstack{{\miniscule (\rule[.5ex]{1.25em}{.18mm})}
  \\ [-.7ex] $#1$}}

\def\squark    {{\ensuremath{\Ps}}\xspace}

\def\cquark    {{\ensuremath{\Pc}}\xspace}

\def\bquark    {{\ensuremath{\Pb}}\xspace}

\def\pion   {{\ensuremath{\Ppi}}\xspace}

\def\pip    {{\ensuremath{\pion^+}}\xspace}
\def\pim    {{\ensuremath{\pion^-}}\xspace}

\def\kaon    {{\ensuremath{\PK}}\xspace}
  \def\Kbar    {{\kern 0.2em\overline{\kern -0.2em \PK}{}}\xspace}

\def\KorKbar    {\kern 0.18em\optbar{\kern -0.18em K}{}\xspace}

\def\Kp      {{\ensuremath{\kaon^+}}\xspace}
\def\Km      {{\ensuremath{\kaon^-}}\xspace}

  \def\Dbar    {{\kern 0.2em\overline{\kern -0.2em \PD}{}}\xspace}

\def\DorDbar    {\kern 0.18em\optbar{\kern -0.18em D}{}\xspace}

\def\B       {{\ensuremath{\PB}}\xspace}
\def\Bbar    {{\ensuremath{\kern 0.18em\overline{\kern -0.18em \PB}{}}}\xspace}

\def\BorBbar    {\kern 0.18em\optbar{\kern -0.18em B}{}\xspace}

\def\Bu      {{\ensuremath{\B^+}}\xspace}

\def\Bp      {{\ensuremath{\Bu}}\xspace}

\def\Bs      {{\ensuremath{\B^0_\squark}}\xspace}

\def\Bc      {{\ensuremath{\B_\cquark^+}}\xspace}

\def\jpsi     {{\ensuremath{{\PJ\mskip -3mu/\mskip -2mu\Ppsi\mskip 2mu}}}\xspace}

\def\chiczero {{\ensuremath{\Pchi_{\cquark 0}}}\xspace}

  \def\Y#1S{\ensuremath{\PUpsilon{(#1S)}}\xspace}

\def\Lbar        {{\ensuremath{\kern 0.1em\overline{\kern -0.1em\PLambda}}}\xspace}
\def\LorLbar    {\kern 0.18em\optbar{\kern -0.18em \PLambda}{}\xspace}

\def\to                 {\ensuremath{\rightarrow}\xspace}

\def\AT#1     {\ensuremath{A_{\mathrm{T}}^{#1}}\xspace}           

\def\C#1      {\ensuremath{\mathcal{C}_{#1}}\xspace}                       
\def\Cp#1     {\ensuremath{\mathcal{C}_{#1}^{'}}\xspace}                    
\def\Ceff#1   {\ensuremath{\mathcal{C}_{#1}^{\mathrm{(eff)}}}\xspace}        
\def\Cpeff#1  {\ensuremath{\mathcal{C}_{#1}^{'\mathrm{(eff)}}}\xspace}       
\def\Ope#1    {\ensuremath{\mathcal{O}_{#1}}\xspace}                       
\def\Opep#1   {\ensuremath{\mathcal{O}_{#1}^{'}}\xspace}                    

\newcommand{\tev}{\ifthenelse{\boolean{inbibliography}}{\ensuremath{~T\kern -0.05em eV}\xspace}{\ensuremath{\mathrm{\,Te\kern -0.1em V}}}\xspace}
\newcommand{\gev}{\ensuremath{\mathrm{\,Ge\kern -0.1em V}}\xspace}
\newcommand{\mev}{\ensuremath{\mathrm{\,Me\kern -0.1em V}}\xspace}
\newcommand{\kev}{\ensuremath{\mathrm{\,ke\kern -0.1em V}}\xspace}
\newcommand{\ev}{\ensuremath{\mathrm{\,e\kern -0.1em V}}\xspace}
\newcommand{\gevc}{\ensuremath{{\mathrm{\,Ge\kern -0.1em V\!/}c}}\xspace}
\newcommand{\mevc}{\ensuremath{{\mathrm{\,Me\kern -0.1em V\!/}c}}\xspace}
\newcommand{\gevcc}{\ensuremath{{\mathrm{\,Ge\kern -0.1em V\!/}c^2}}\xspace}
\newcommand{\gevgevcccc}{\ensuremath{{\mathrm{\,Ge\kern -0.1em V^2\!/}c^4}}\xspace}
\newcommand{\mevcc}{\ensuremath{{\mathrm{\,Me\kern -0.1em V\!/}c^2}}\xspace}

\def\invfb   {\ensuremath{\mbox{\,fb}^{-1}}\xspace}

\def\ps   {\ensuremath{{\rm \,ps}}\xspace}

\def\gsim{{~\raise.15em\hbox{$>$}\kern-.85em
          \lower.35em\hbox{$\sim$}~}\xspace}
\def\lsim{{~\raise.15em\hbox{$<$}\kern-.85em
          \lower.35em\hbox{$\sim$}~}\xspace}

\def\pt         {\mbox{$p_{\rm T}$}\xspace}

\def\tell1  {TELL1\xspace}
\def\ukl1   {UKL1\xspace}

\newcommand{\eg}{\mbox{\itshape e.g.}\xspace}

\newcommand{\vs}{\mbox{\itshape vs.}\xspace}

%% file: bhhh-symbols-def.tex
\def\kkpiplus {\ensuremath{\Bp \to \Kp \Km \pip}\xspace}

\def\bckkpiplus {\ensuremath{\Bc \to \Kp \Km \pip}\xspace}



%% file: title-LHCb-PAPER.tex

\begin{titlepage}
\pagenumbering{roman}

\vspace*{-1.5cm}
\centerline{\large EUROPEAN ORGANIZATION FOR NUCLEAR RESEARCH (CERN)}
\vspace*{1.5cm}
\hspace*{-0.5cm}
\begin{tabular*}{\linewidth}{lc@{\extracolsep{\fill}}r}
\ifthenelse{\boolean{pdflatex}}
{\vspace*{-2.7cm}\mbox{\!\!\!\includegraphics[width=.14\textwidth]{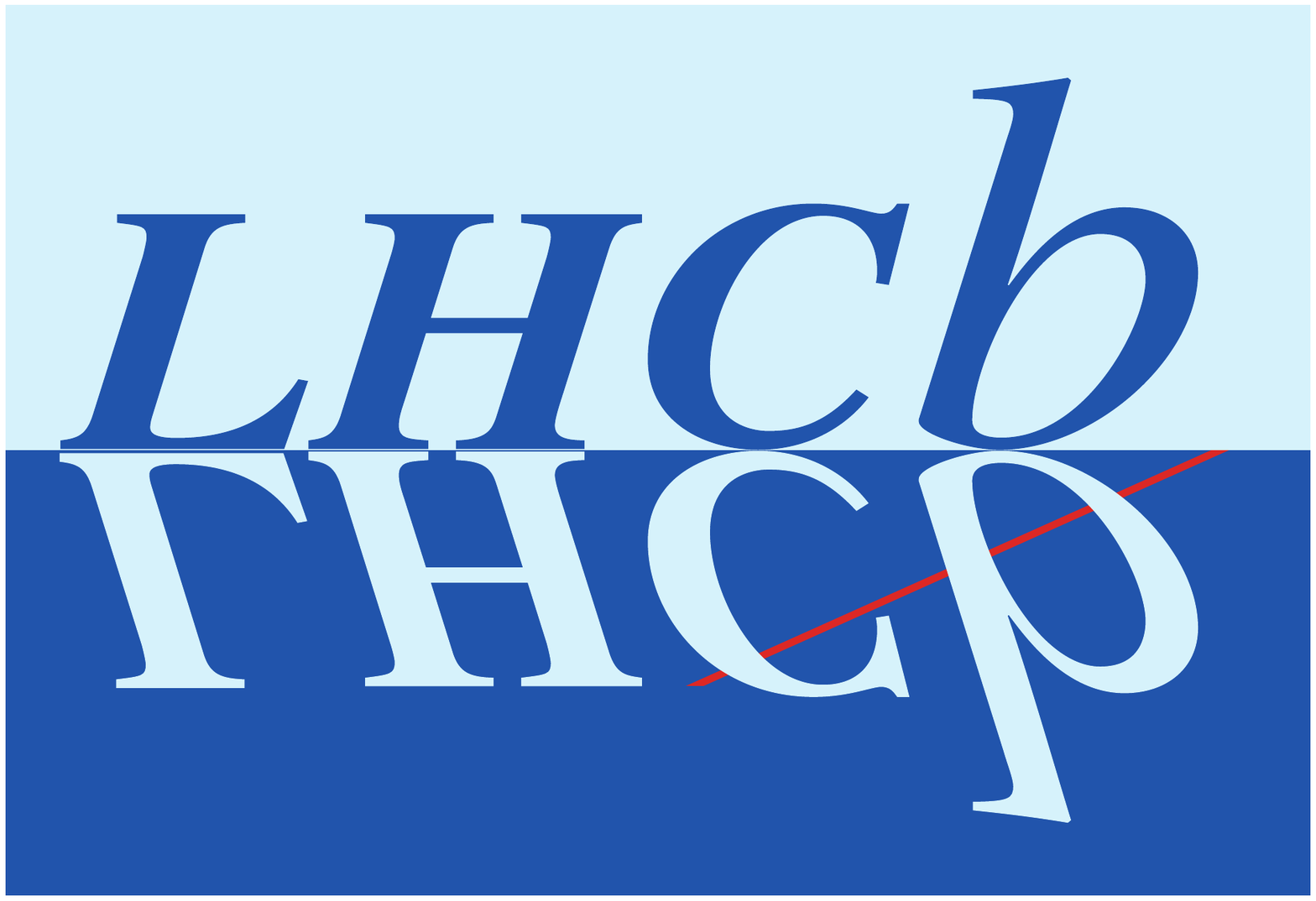}} & &}%
{\vspace*{-1.2cm}\mbox{\!\!\!\includegraphics[width=.12\textwidth]{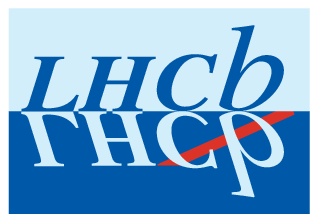}} & &}%
\\
 & & CERN-EP-2016-176 \\  
 & & LHCb-PAPER-2016-022 \\  
 & & July 20, 2016 \\ 
\end{tabular*}

\vspace*{2.0cm}

{\bf\boldmath\huge
\begin{center}
  Study of \Bc decays to the $\Kp\Km\pip$ final state and evidence for the decay $\Bc\to\chiczero\pip$
\end{center}
}

\vspace*{1.0cm}

\begin{center}
The LHCb collaboration\footnote{Authors are listed at the end of this article.}
\end{center}

\vspace{\fill}

\begin{abstract}
  \noindent
  A study of \bckkpiplus decays is performed for the first time using data corresponding to an integrated luminosity of 3.0\invfb collected by the \lhcb experiment in $pp$ collisions at centre-of-mass energies of $7$ and $8$\tev. Evidence for the decay $\Bc\to\chiczero(\to\Kp\Km)\pip$ is reported with a significance of 4.0 standard deviations, giving $\frac{\sigma(\Bc)}{\sigma(B^+)}\times\mathcal{B}(\Bc\to\chiczero\pip)=(9.8^{+3.4}_{-3.0}(\mathrm{stat})\pm 0.8(\mathrm{syst}))\times 10^{-6}$. Here $\mathcal{B}$ denotes a branching fraction while $\sigma(\Bc)$ and $\sigma(B^+)$ are the production cross-sections for \Bc and \Bp mesons. An indication of $\overline b c$ weak annihilation is found for the region $m(\Km\pip)<1.834\gevcc$, with a significance of 2.4 standard deviations. 
\end{abstract}

\vspace*{1.0cm}

\begin{center}
  Published in Phys. Rev. D94 (2016) 091102(R)
\end{center}

\vspace{\fill}

{\footnotesize 
\centerline{\copyright~CERN on behalf of the \lhcb collaboration, license \href{http://creativecommons.org/licenses/by/4.0/}{CC-BY-4.0}.}}
\vspace*{2mm}

\end{titlepage}


\newpage
\setcounter{page}{2}
\mbox{~}


\cleardoublepage

%% file: introduction.tex
Heavy flavour physics involves studying the decays of hadrons containing at least one \bquark or \cquark valence quark, with the possibility of making precision measurements of Standard Model (SM) parameters and detecting effects of new physics. The \Bc meson ($\overline b c$), the only currently established hadron having two different heavy-flavour quarks, has the particularity of decaying weakly through either of its flavours.\footnote{Charge conjugation is implied throughout the paper.} In the SM, the \Bc decays with no charm and beauty particles in the final or intermediate states can proceed only via  $\overline b c\to W^+\to u\overline q$ ($q=d,s$) annihilation, with an amplitude proportional to the product of CKM matrix elements $V_{cb}^*V_{uq}$. Calculations predict branching fractions in the range $10^{-8}-10^{-6}$\cite{LHCb_BcPred,XiaoLiuLu,XiaoAndLiu}. Any significant enhancement could indicate the presence of $\overline b c$ annihilations involving particles beyond the SM, such as a mediating charged Higgs boson (see \eg Ref.~\cite{Hou_Lepton,Extended_Higgs}). 

Experimentally, the decays of \Bc mesons to three light charged hadrons provide a good way to study such processes. These decay modes have a large available phase space and can include other processes such as $\Bc\to D^0(\to K\pi)h^+$ ($h=\pi,~K$) \cite{BctoDH_Th} mediated by $\overline b\to \overline u$ and $\overline b \to\overline d,\overline s$ transitions, $\Bc\to B^0_q(\to h_1^+ h_2^-) h_3^+$ decays \cite{BctoBh_Th} mediated by $c\to q$ transitions, or charmonium modes $\Bc\to [c\overline c](\to h_1^+h_1^-)h_2^+$ \cite{BcJpsiPi_Th} mediated by the $b\to c$ transition \cite{Suppl_Material}.
In this study, special consideration is given to decays leading to a $\Kp\Km\pip$ final state in the region well below the $D^0$ mass, taken to be $m(\Km\pip)<1.834\gevcc$, where, after removing possible contributions from $([c\overline c],\Bs)\to\Kp\Km$, only the annihilation process remains. The other contributions listed above are also examined. The $\Bp\to {\overline D}^0(\to\Kp\Km)\pi^+$ decay is used as a normalization mode to derive the quantity
\begin{equation}
R_f \equiv \frac{\sigma(\Bc)}{\sigma(B^+)}\times\mathcal{B}(\Bc\to f),
\end{equation}
where $\mathcal{B}$ is the branching fraction, and $\sigma(\Bc)$ and $\sigma(B^+)$ are the production cross-sections of the $\Bc$ and $\Bp$ mesons. The quantity $R_f$ is measured in the fiducial region $\pt(B)<20\gevc$ and $2.0<y(B)<4.5$, where \pt is the component of the momentum transverse to the proton beam and $y$ denotes the rapidity. The data sample used corresponds to integrated luminosities of 1.0 and 2.0\invfb collected by the \lhcb experiment at $7$ and $8$ \tev centre-of-mass energies in $pp$ collisions, respectively. Since the kinematics of $B$ meson production is very similar at the two energies, the ratio $\frac{\sigma(\Bc)}{\sigma(B^+)}$ is assumed to be the same for all the measurements discussed in this Letter.

%% file: detector_short.tex
The \lhcb detector is a single-arm forward spectrometer covering the \mbox{pseudorapidity} range $2<\eta <5$, described in detail in Ref.~\cite{Alves:2008zz,LHCb-DP-2014-002}. The detector allows the reconstruction of both charged and neutral particles. For this analysis, the ring-imaging Cherenkov (RICH) detectors \cite{RichPerf}, distinguishing pions, kaons and protons, are particularly important. Simulated events are produced using the software described in Refs.~\cite{Sjostrand:2006za,*Sjostrand:2007gs,LHCb-PROC-2010-056, Lange:2001uf, Allison:2006ve, *Agostinelli:2002hh, LHCb-PROC-2011-006,BcVegPy,BcVegPy2}.

%% file: selection_short.tex
The $B_{(c)}^+\to\Kp\Km\pip$ decay candidates are reconstructed applying the same selection procedure as in Ref.~\cite{LHCb-PAPER-2016-001}.
A similar multivariate analysis is implemented, using a boosted decision tree (BDT) classifier\cite{BDT_theory}. Particle identification (PID) requirements are then applied to reduce the combinatorial background and suppress the cross-feed from pions misidentified as kaons. The BDT and PID requirements are optimized to maximize the sensitivity to small event yields. 

The \Bc signal yield is determined from a simultaneous fit in three bins of the BDT output $\mathcal{O}_{\mathrm{BDT}}$, $0.04<\mathcal{O}_{\mathrm{BDT}}<0.12$, $0.12<\mathcal{O}_{\mathrm{BDT}}<0.18$ and $\mathcal{O}_{\mathrm{BDT}}>0.18$, each having similar expected yield but different levels of background \cite{LHCb-PAPER-2016-001}.
The normalization channel $\Bp\to {\overline D}^0(\to\Kp\Km)\pi^+$ uses the same BDT classifier, with tighter PID requirements to suppress the abundant background from $\Bp\to\Kp\pim\pip$ decays. Its yield is determined requiring $\mathcal{O}_{\mathrm{BDT}}>0.04$, and demanding $1.834<m(\Kp\Km)<1.894\gevcc$ to remove charmless $\Bp\to\Kp\Km\pip$ candidates.

%% file: fits.tex
\begin{figure}[t]
\begin{center}
  \includegraphics[width=0.5\linewidth,keepaspectratio=true]{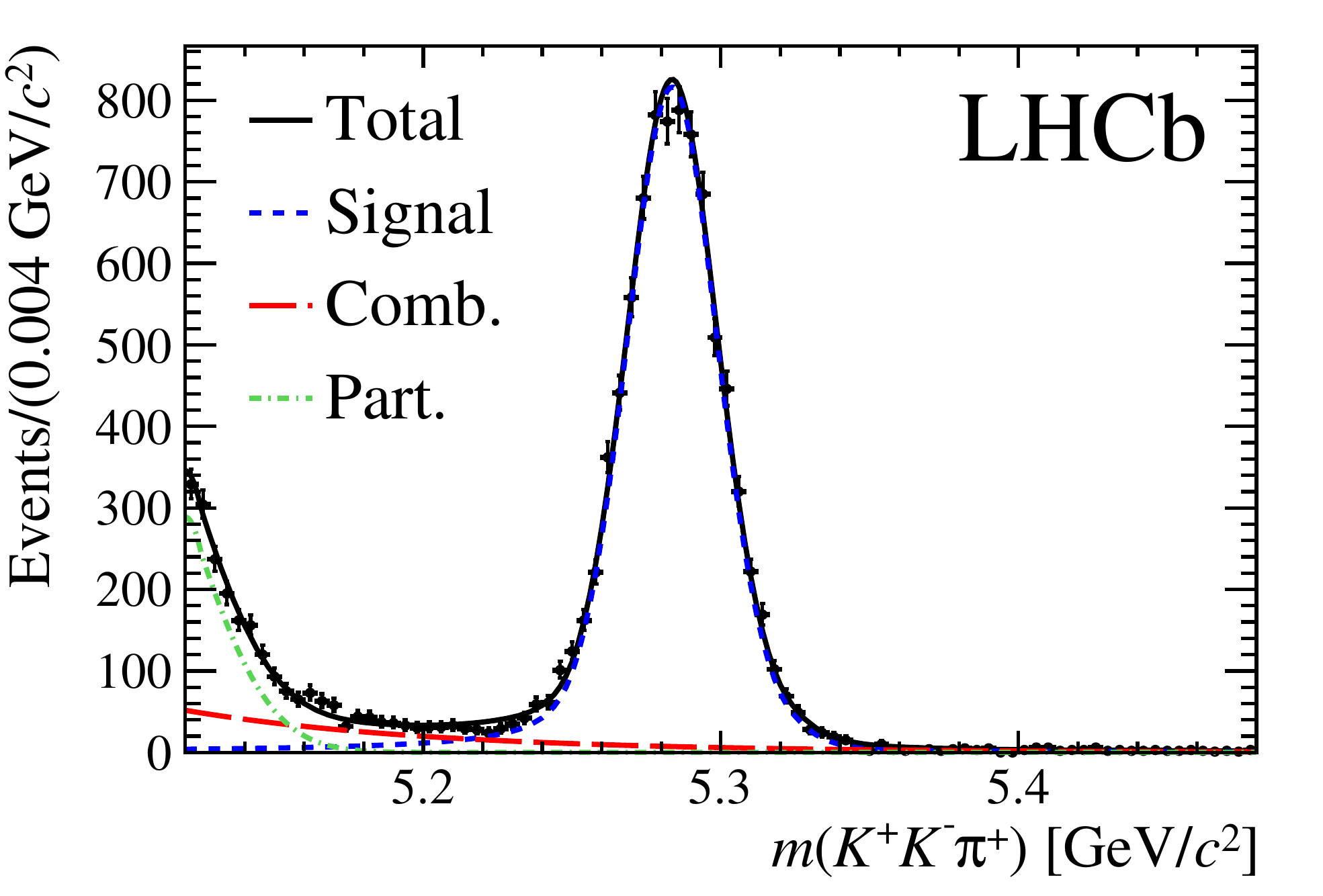}
\end{center}
  \caption{Fit to the $\Kp\Km\pip$ invariant mass for the \Bp candidates, with $1.834<m(\Kp\Km)<1.894\gevcc$. The contributions from the signal $\Bp\to {\overline D}^0(\to\Kp\Km)\pip$, combinatorial background (Comb.) and partially reconstructed background (Part.) obtained from the fit are shown.\label{fig:BuD0KKPi_fit}}
\end{figure}

 Signal and background yields are obtained from extended unbinned maximum likelihood fits to the distribution of the invariant mass of the $\Kp\Km\pip$ combinations. The \bckkpiplus and \kkpiplus signals are each modelled by the sum of two Crystal Ball functions \cite{CB_shape} with a common mean. For \bckkpiplus all the shape parameters and the relative yields in each bin of $\mathcal{O}_{\mathrm{BDT}}$ are fixed to the values obtained in the simulation, while for \kkpiplus the mean and the core width are allowed to vary freely in the fit.
A Fermi-Dirac function is used to model a possible partially reconstructed component from decays with $\Kp\Km\pip\pi^0$ final states where the neutral pion is not reconstructed, resulting in a $\Kp\Km\pip$ invariant mass below the nominal $\Bc$ or $\Bp$ mass. All shape parameters of these background components are fixed to the values obtained from simulation. The combinatorial background is modelled by an exponential function.
Figure \ref{fig:BuD0KKPi_fit} shows the result of the fit to determine the yield of the $\Bp\to {\overline D}^0(\to\Kp\Km)\pip$ channel, $N_u=8577\pm109$.

\begin{figure}[t]
\begin{center}
  \includegraphics[width=0.5\linewidth,keepaspectratio=true]{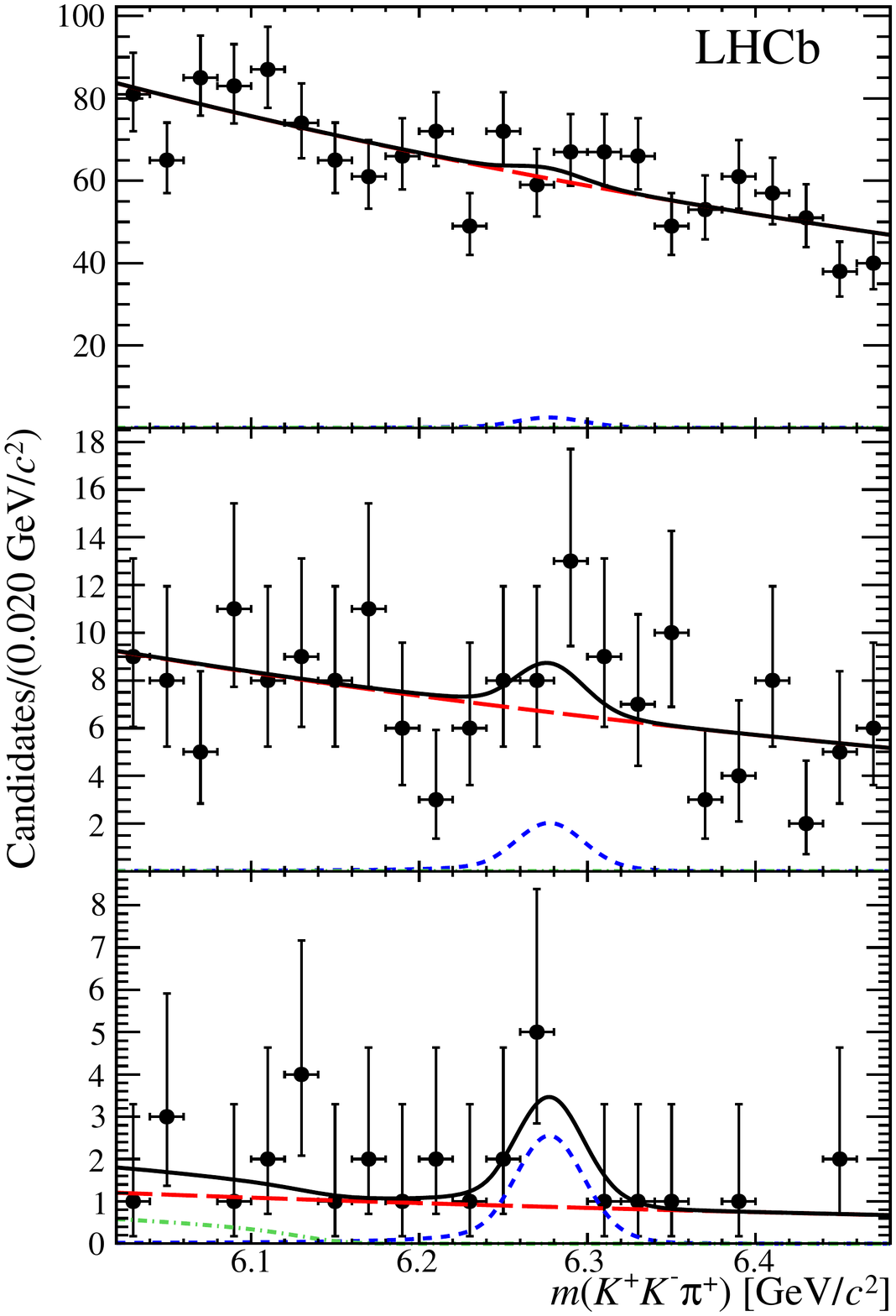}
\end{center}
  \caption{Projection of the fit to the $\Kp\Km\pip$ invariant mass in the \Bc region, in the bins of BDT output used in the analysis: (top) $0.04<\mathcal{O}_{\mathrm{BDT}}<0.12$, (middle) $0.12<\mathcal{O}_{\mathrm{BDT}}<0.18$ and (bottom) $\mathcal{O}_{\mathrm{BDT}}>0.18$, for $m(\Km\pip)<1.834\gevcc$, including the vetoes in $m(\Kp\Km)$ (see text). Apart from the signal type, which is given by $\Bc\to\Kp\Km\pip$, the contributions are indicated according to the same scheme as in Fig.~\ref{fig:BuD0KKPi_fit}.\label{fig:Bc_An_fit}}
\end{figure}

In the \Bc region $6.0<m(\Kp\Km\pip)<6.5\gevcc$, the signals are fitted separately for regions of the phase space corresponding to the different expected contributions: the annihilation region ($m(\Km\pip)<1.834\gevcc$), the $D^0\to\Km\pip$ region ($1.834<m(\Km\pip)<1.894\gevcc$), and the $\Bs\to\Km\Kp$ region ($5.3<m(\Kp\Km)<5.4\gevcc$). For the first two regions, the ranges $3.38<m(\Kp\Km)<3.46\gevcc$ and $5.2<m(\Kp\Km)<5.5\gevcc$ are vetoed to remove contributions from \chiczero (as discussed below) and $B^0_{(s)}\to h_1^+ h_2^-$ decays. A possible signal is seen in the annihilation region, as shown in Fig.~\ref{fig:Bc_An_fit}. The corresponding yield is $N_c=20.8^{+11.4}_{-9.9}$, with a statistical significance of 2.5 standard deviations ($\upsigma$), inferred from the difference in the logarithm of the likelihood for fits with and without the signal component.

\begin{figure}[t]
\begin{center}
  \includegraphics[width=0.49\linewidth,keepaspectratio=true]{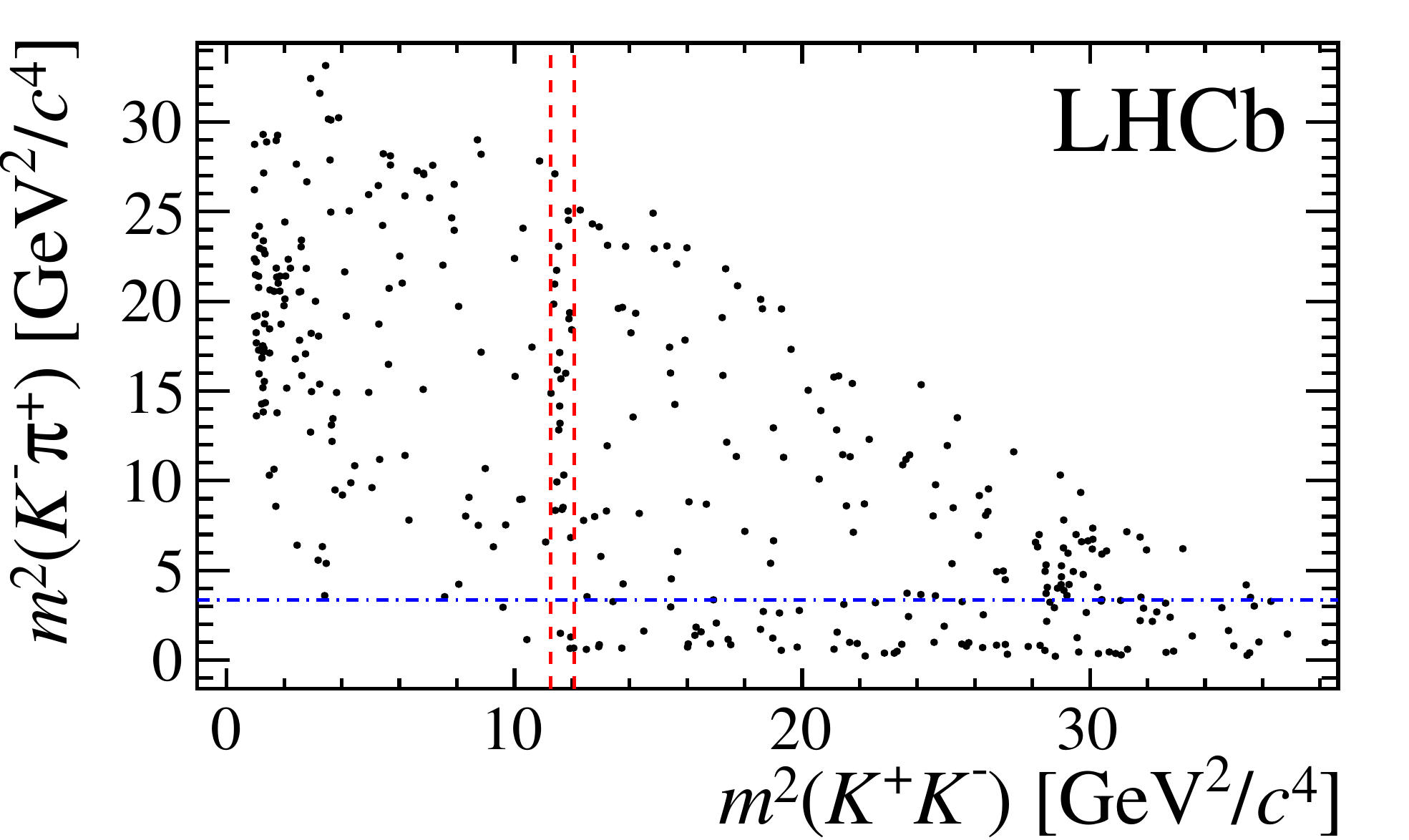}
  \includegraphics[width=0.49\linewidth,keepaspectratio=true]{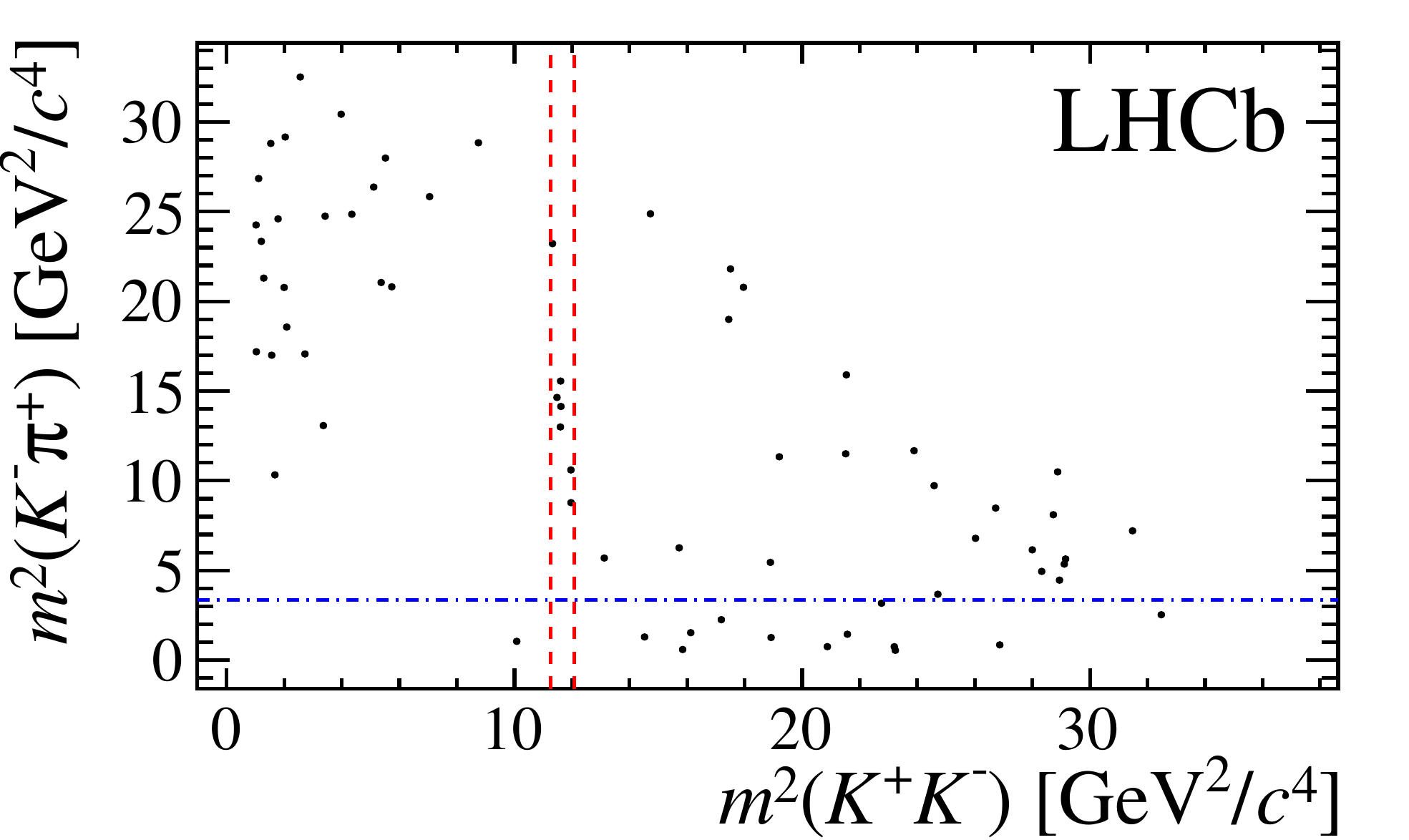}
\end{center}
  \caption{Distribution of events for the signal region $6.2<m(\Kp\Km\pip)<6.35\gevcc$ in the $m^2(\Km\pip)$ \vs $m^2(\Kp\Km)$ plane for (left) $\mathcal{O}_{\mathrm{BDT}}>0.12$ and (right) $\mathcal{O}_{\mathrm{BDT}}>0.18$. The vertical red dashed lines represent a band of width $\pm60\mevcc$ around the \chiczero mass. The horizontal blue dot-dashed line indicates the upper bound of the annihilation region at $m(\Km\pip)=1.834\gevcc$, representing 17\% of the available phase space area.\label{fig:DalitzBc_sigreg}}
\end{figure}

The distribution of events in the $m^2(\Km\pip)$ \vs $m^2(\Kp\Km)$ plane, for the \Bc signal region $6.2<m(\Kp\Km\pip)<6.35\gevcc$, is shown in Fig.~\ref{fig:DalitzBc_sigreg}. A concentration of events is observed around $m^2(\Kp\Km)\sim 11\gevgevcccc$. A one-dimensional projection of $m(\Kp\Km)$ shows clustering near $3.41\gevcc$, close to the mass of the charmonium state \chiczero. Among all the charmonia, \chiczero has the highest branching fraction into the $\Kp\Km$ final state \cite{PDG2014}. The accumulation of events near $m^2(\Kp\Km)\sim 29\gevgevcccc$ for the loose $\mathcal{O}_{\mathrm{BDT}}$ cut appears to be mainly caused by $\Bs\to\Kp\Km$ decays combined with random pions since no peak is seen in $m(\Kp\Km\pip)$ at the \Bc mass \cite{Suppl_Material}.

\begin{figure}[t]
\begin{center}
  \includegraphics[width=0.49\linewidth,keepaspectratio=true]{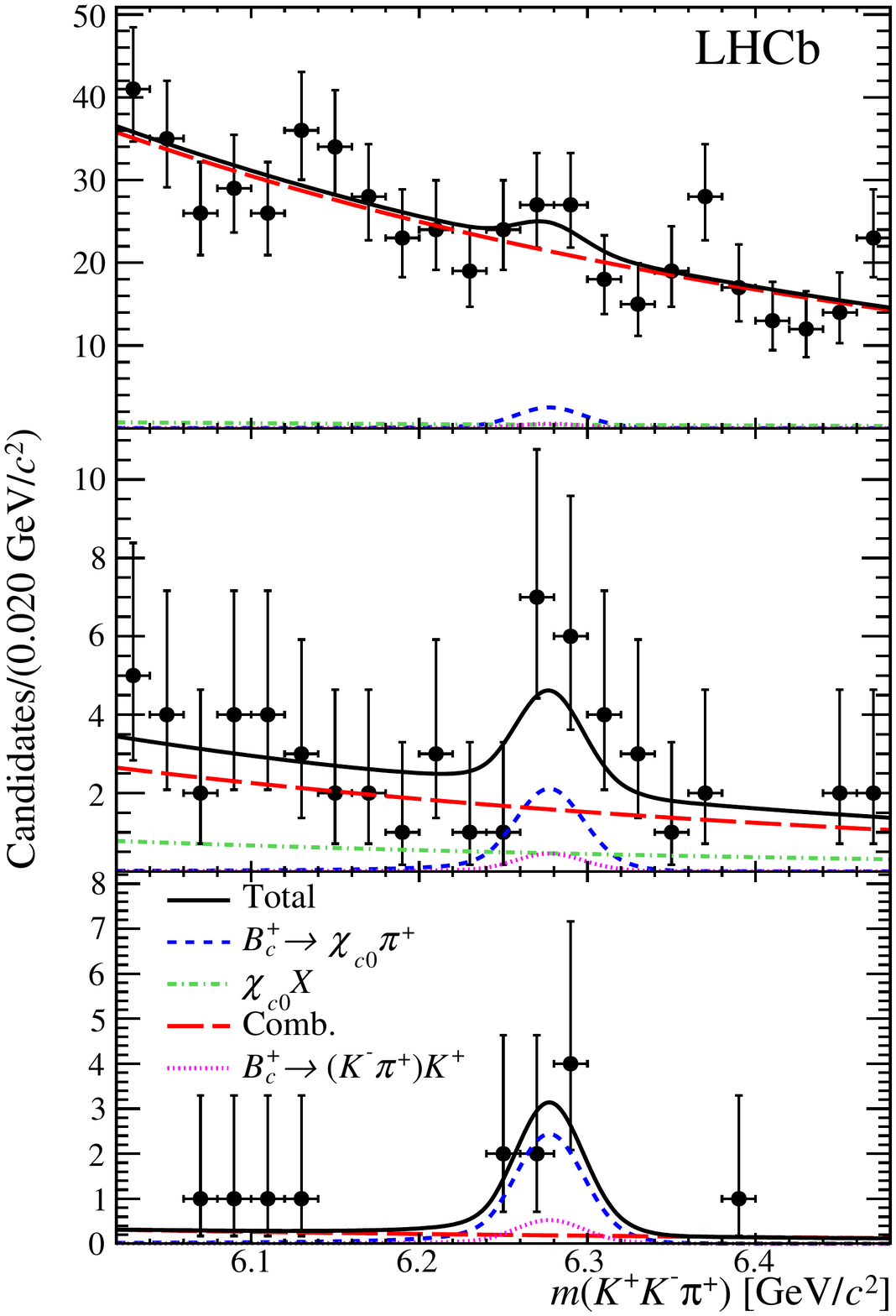}
  \includegraphics[width=0.49\linewidth,keepaspectratio=true]{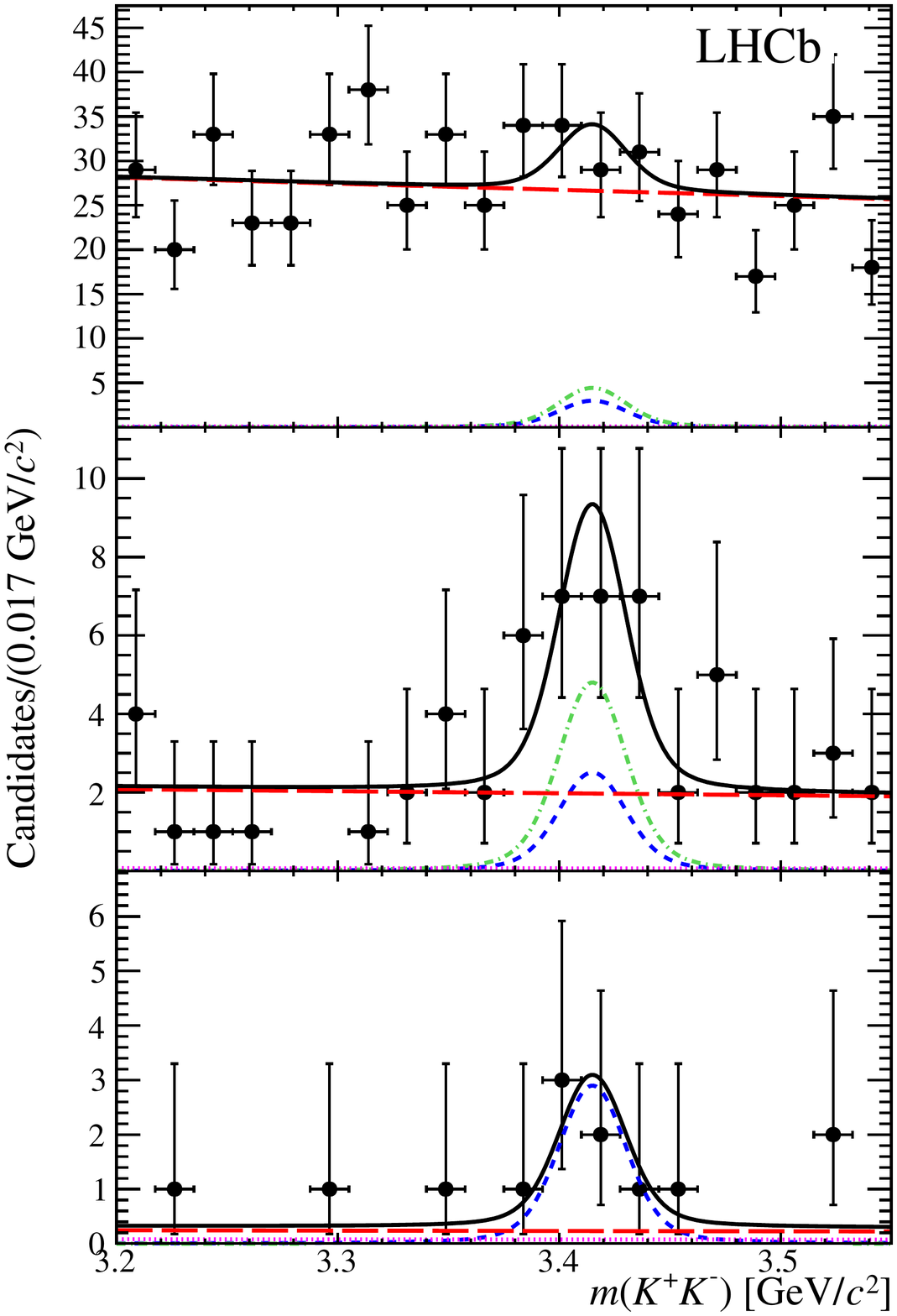}
\end{center}
  \caption{Fit projections to the (left) $\Kp\Km\pip$ and (right) $\Kp\Km$ invariant masses, in the bins of BDT output (top) $0.04<\mathcal{O}_{\mathrm{BDT}}<0.12$, (middle) $0.12<\mathcal{O}_{\mathrm{BDT}}<0.18$ and (bottom) $\mathcal{O}_{\mathrm{BDT}}>0.18$, for the extraction of the $\Bc\to\chiczero(\to\Kp\Km)\pip$ signal. The contributions from the $\Bc\to \chiczero(\to\Kp\Km)\pip$ signal, combinatorial background (Comb.), possible pollution from the annihilation region $\Bc\to (\Km\pip)\Kp$, and combinations of $\chiczero\to\Kp\Km$ with a random track $X$ are shown.\label{fig:Bc_Chic0Pi_fit}}
\end{figure}

To determine the $\Bc\to\chiczero(\to\Kp\Km)\pip$ signal yield, the two-dimensional $m(\Kp\Km\pip)$ \vs $m(\Kp\Km)$ distributions are fitted simultaneously for each of the three BDT bins. The $m(\Kp\Km\pip)$ distribution is modelled in the same way as described above. The $m(\Kp\Km)$ distribution is fitted in the range $3.20<m(\Kp\Km)<3.55\gevcc$. The $\chiczero\to\Kp\Km$ shape is modelled by a Breit--Wigner function, with mean and width fixed to their known values~\cite{PDG2014}, convolved with a Gaussian resolution function, while a first-order polynomial is used to represent the $\Kp\Km$ background. Figure \ref{fig:Bc_Chic0Pi_fit} shows the projections of the fit result. The yield obtained is $N_\chiczero=20.8^{+7.2}_{-6.4}$, with a statistical significance of 4.1 $\upsigma$.
The fits for the $D^0$ and \Bs regions, where no signal is observed, can be found at Ref.~\cite{Suppl_Material}.

%% file: eff_short.tex
For each region of phase-space considered, the efficiencies for the signals, $\epsilon_c$, and normalization channel, $\epsilon_u$, are inferred from simulated samples and are corrected using data-driven methods as described in Ref.~\cite{LHCb-PAPER-2016-001}. They include the effects of reconstruction, selection and detector acceptance.
An efficiency map defined in the  $m^2(\Km\pip)$ \vs $m^2(\Kp\Km)$ plane is computed. Because of limited statistics, the distribution of the signal events in the annihilation region is not well known.  Therefore, the efficiency for the annihilation region is estimated in two ways: first, by taking the simple average efficiency from the map for $m(\Km\pip)<1.834\gevcc$ and alternatively, by taking the efficiency weighted according to the sparse distribution of candidates in data in the $m^2(\Km\pip)$ \vs $m^2(\Kp\Km)$ plane. The average of the two values is taken as the efficiency and the difference is treated as a systematic uncertainty (labelled as ``event distribution'' in Table \ref{Tab:systematics}). A correction accounting for the vetoed $m(\Kp\Km)$ regions described above is included. In the calculation of the observable $R_f$ the efficiency ratio $\epsilon_u/\epsilon_c$ is required. The values obtained are $1.698\pm0.015$ for the annihilation region and $1.241\pm0.012$ for the $\Bc\to\chiczero(\Kp\Km)\pip$ mode. The uncertainties are due to the limited sizes of the simulated samples. The differences between the \Bp and \Bc efficiencies are caused by the different lifetimes and masses of the two mesons. 

%% file: results.tex
The measured quantities are determined as
\begin{equation}
\nonumber R_{\mathrm{an},KK\pi}=\frac{N_{c}}{N_{u}}\ \times \frac{\epsilon_{u}}{\epsilon_{c}(\mathrm{an},KK\pi)}\times \mathcal{B}(B^{\pm} \to D^0\pi^{\pm})\times \mathcal{B}(D^0 \to \Kp\Km)
\end{equation}
for the annihilation region, and
\begin{equation}
\nonumber R_{\chiczero\pi}=\frac{\sigma(\Bc)}{\sigma(B^+)}\times\mathcal{B}(\Bc\to\chiczero\pip)=\frac{N_{\chiczero}}{N_{u}}\ \times \frac{\epsilon_{u}}{\epsilon_c(\chiczero)}\times \frac{\mathcal{B}(B^{\pm} \to D^0\pi^{\pm})\times \mathcal{B}(D^0 \to \Kp\Km)}{\mathcal{B}( \chiczero\to \Kp\Km)}
\end{equation} 
for the $\Bc\to\chiczero\pip$ decay, where $\epsilon_{x}$ are the efficiencies and $N_{x}$ are the yields obtained from the fits.

Systematic uncertainties are associated with the yield ratios, the efficiency ratios and the branching fractions $\mathcal{B}(\Bp \to \overline{D}^{0}\pip)=(4.81\pm0.15)\times10^{-3}$, $\mathcal{B}(D^{0}\to\Km\Kp)=(4.01\pm0.07)\times10^{-3}$ and $\mathcal{B}(\chiczero\to\Km\Kp)=(5.91\pm0.32)\times10^{-3}$~\cite{PDG2014}. Table~\ref{Tab:systematics} summarizes the uncertainties. The yields are affected by the uncertainties on the fit functions and parameters, and by the variation of the yield fractions in the BDT output bins, due to the uncertainty on the BDT output distribution. The uncertainties on the efficiency ratios reflect the PID calibration, the limited sizes of the simulated samples, the effect of the detector acceptance, the \Bc lifetime $0.507\pm0.009\ps$ \cite{HFAG}, and the trigger and fiducial cut corrections.

\begin{table}[tb]
  \begin{center}
\caption{Relative systematic uncertainties (in \%) of the measurements of $R_{\mathrm{an},KK\pi}$ and $R_{\chiczero\pi}$.}\label{Tab:systematics}
\begin{tabular}{lcc}	
\centering
 Source & $R_{\mathrm{an},KK\pi}$ & $R_{\chiczero\pi}$\\ 
\hline 
Normalisation yield & 1.3 & 1.3 \\
Event distribution & 1.6 & -- \\
Fit model & 2.4 & 2.3\\ 
BDT shape & 5.0 & 2.9\\
PID & 1.0 & 1.0\\
Simulation & 0.8 & 0.8\\
Detector acceptance & 0.4 & 0.3\\
\Bc lifetime & 2.0 & 2.0\\
Hardware trigger & 1.5 & 1.4\\
Fiducial cut & 0.1 & 0.1\\
Branching fractions & 3.6 & 6.2\\
\hline
Total & 7.5 & 7.8\\
\end{tabular}
\end{center}    
\end{table}

The results obtained are $R_{\mathrm{an},KK\pi}=(8.0^{+4.4}_{-3.8}(\mathrm{stat})\pm 0.6(\mathrm{syst}))\times 10^{-8}$ and $R_{\chiczero\pi}=(9.8^{+3.4}_{-3.0}(\mathrm{stat})\pm 0.8(\mathrm{syst}))\times 10^{-6}$. Accounting for the systematic uncertainties related to the signal extraction, the significances of these measurements are 2.4 $\upsigma$ and 4.0 $\upsigma$, respectively. For the annihilation region, a 90(95)\% confidence level (CL) upper limit, $R_{\mathrm{an},KK\pi}<15(17)\times 10^{-8}$, is estimated by making a scan of $R_{\mathrm{an},KK\pi}$, comparing profile likelihood ratios for the ``signal+background'' and ``background-only'' hypotheses \cite{CL-art,Suppl_Material}.

For the modes $\Bc\to\Bs(\to\Kp\Km)\pip$ and $\Bc\to D^0(\to\Km\pip)\Kp$, no significant deviation from the background-only hypothesis is observed. Using $\mathcal{B}(\Bs\to\Kp\Km)=(2.50\pm0.17)\times 10^{-5}$ and $\mathcal{B}(D^0\to\Km\pip)=(3.93\pm0.04)\%$~\cite{PDG2014}, the following 90(95)\% CL upper limits are obtained: $R_{\Bs\pi}\equiv\frac{\sigma(\Bc)}{\sigma(B^+)}\times\mathcal{B}(B_{c}^{+} \to \Bs\pi^{+})<4.5(5.4)\times 10^{-3}$ and $R_{D^0K}\equiv\frac{\sigma(\Bc)}{\sigma(B^+)}\times\mathcal{B}(B_{c}^{+} \to D^0K^+)<1.3(1.6)\times 10^{-6}$. The first limit is consistent with the result of Ref.~\cite{LHCb-PAPER-2013-044}, which gives $R_{\Bs\pi}=(6.2\pm1.0)\times10^{-4}$, using $\sigma(\Bs)/\sigma(B^{+})=0.258\pm0.016$ \cite{LHCb-PAPER-2011-018,LHCb-PAPER-2012-037}.

In summary, a study of \Bc meson decays to the $\Kp\Km\pip$ final state has been performed in the fiducial region $\pt(B)<20\gevc$ and $2.0<y(B)<4.5$. 
Evidence for the decay $\Bc\to\chiczero\pip$ is found at 4.0 $\upsigma$ significance.
This result can be compared to the measurement involving another charmonium mode, $\frac{\sigma(\Bc)}{\sigma(B^+)}\times\mathcal{B}(\Bc\to \jpsi\pip)=(7.0\pm0.3)\times10^{-6}$, obtained from Refs.~\cite{LHCb-PAPER-2014-050,PDG2014}.

A indication of $\overline b c$ weak annihilation with a significance of 2.4 $\upsigma$ is reported in the region $m(\Km\pip)<1.834\gevcc$. The branching fraction of $\Bc\to \overline{K}^{*0}(892)\Kp$ has been recently predicted to be $(10.0^{+1.8}_{-3.4})\times 10^{-7}$~\cite{XiaoAndLiu}. The contribution of the mode $\Bc\to \overline{K}^{*0}(892)(\to \Km\pip)\Kp$ to $R_{\mathrm{an},KK\pi}$ could be prominent, for which an estimate is made as follows.
Using the predictions listed in Ref.\cite{BcJpsiPi_Th2} for $\mathcal{B}(\Bc\to \jpsi\pip)$, which span the range $[0.34,2.9]\times10^{-3}$, and the value of $\frac{\sigma(\Bc)}{\sigma(B^+)}\times\mathcal{B}(\Bc\to \jpsi\pip)$ based on Ref.~\cite{LHCb-PAPER-2014-050} quoted above, $\frac{\sigma(\Bc)}{\sigma(B^+)}\sim[0.23,2.1]\%$ is obtained. Combined with the prediction of Ref.~\cite{XiaoAndLiu}, a value of $\frac{\sigma(\Bc)}{\sigma(B^+)}\times\mathcal{B}(\Bc\to \overline{K}^{*0}(892)(\to \Km\pip)\Kp)\sim[0.1,1.7]\times 10^{-8}$ is obtained, including the theoretical uncertainties and the $\overline{K}^{*0}(892)\to \Km\pip$ branching fraction. This estimate is lower than the $R_{\mathrm{an},KK\pi}$ measurement. The statistical uncertainty, however, is at present too large to make a definite statement. The data being accumulated in the current run of the LHC will allow \lhcb to clarify whether the weak annihilation process of \Bc meson decays involves significant contributions from heavier $\Km\pip$ states, or is enhanced by other sources.

%% file: acknowledgements.tex
\section*{Acknowledgements}

\noindent We express our gratitude to our colleagues in the CERN
accelerator departments for the excellent performance of the LHC. We
thank the technical and administrative staff at the LHCb
institutes. We acknowledge support from CERN and from the national
agencies: CAPES, CNPq, FAPERJ and FINEP (Brazil); NSFC (China);
CNRS/IN2P3 (France); BMBF, DFG and MPG (Germany); INFN (Italy); 
FOM and NWO (The Netherlands); MNiSW and NCN (Poland); MEN/IFA (Romania); 
MinES and FANO (Russia); MinECo (Spain); SNSF and SER (Switzerland); 
NASU (Ukraine); STFC (United Kingdom); NSF (USA).
We acknowledge the computing resources that are provided by CERN, IN2P3 (France), KIT and DESY (Germany), INFN (Italy), SURF (The Netherlands), PIC (Spain), GridPP (United Kingdom), RRCKI and Yandex LLC (Russia), CSCS (Switzerland), IFIN-HH (Romania), CBPF (Brazil), PL-GRID (Poland) and OSC (USA). We are indebted to the communities behind the multiple open 
source software packages on which we depend.
Individual groups or members have received support from AvH Foundation (Germany),
EPLANET, Marie Sk\l{}odowska-Curie Actions and ERC (European Union), 
Conseil G\'{e}n\'{e}ral de Haute-Savoie, Labex ENIGMASS and OCEVU, 
R\'{e}gion Auvergne (France), RFBR and Yandex LLC (Russia), GVA, XuntaGal and GENCAT (Spain), Herchel Smith Fund, The Royal Society, Royal Commission for the Exhibition of 1851 and the Leverhulme Trust (United Kingdom).

%% file: LHCb-PAPER-2016-022.bbl
\ifx\mcitethebibliography\mciteundefinedmacro
\PackageError{LHCb.bst}{mciteplus.sty has not been loaded}
{This bibstyle requires the use of the mciteplus package.}\fi
\providecommand{\href}[2]{#2}

%% file: LHCb_Authorship_flat_01-Jun-2016.tex
\centerline{\large\bf LHCb collaboration}
\begin{flushleft}
\small
R.~Aaij$^{39}$,
B.~Adeva$^{38}$,
M.~Adinolfi$^{47}$,
Z.~Ajaltouni$^{5}$,
S.~Akar$^{6}$,
J.~Albrecht$^{10}$,
F.~Alessio$^{39}$,
M.~Alexander$^{52}$,
S.~Ali$^{42}$,
G.~Alkhazov$^{31}$
% ...

{\footnotesize \it
% ...
$ ^{5}$Clermont Universit{\'e}, Universit{\'e} Blaise Pascal, CNRS/IN2P3, LPC, Clermont-Ferrand, France\\
$ ^{6}$CPPM, Aix-Marseille Universit{\'e}, CNRS/IN2P3, Marseille, France\\
% ...
$ ^{10}$Fakult{\"a}t Physik, Technische Universit{\"a}t Dortmund, Dortmund, Germany\\
% ...
$ ^{31}$Petersburg Nuclear Physics Institute (PNPI), Gatchina, Russia\\
% ...
$ ^{38}$Universidad de Santiago de Compostela, Santiago de Compostela, Spain\\
$ ^{39}$European Organization for Nuclear Research (CERN), Geneva, Switzerland\\
% ...
$ ^{42}$Nikhef National Institute for Subatomic Physics, Amsterdam, The Netherlands\\
% ...
$ ^{47}$H.H. Wills Physics Laboratory, University of Bristol, Bristol, United Kingdom\\
% ...
$ ^{52}$School of Physics and Astronomy, University of Glasgow, Glasgow, United Kingdom\\
% ...
}
\end{flushleft}